\documentclass[a4paper,fleqn]{cas-dc}

\usepackage[numbers]{natbib}
\usepackage{amsmath}
\usepackage{color}
\usepackage{algorithmic}
\usepackage{array}

\usepackage{fixltx2e}
\usepackage{stfloats}
\usepackage{float}
\usepackage{textcomp}
\usepackage{multirow}

\usepackage[font=small,labelfont=bf,labelsep=none]{caption}
\def\tsc#1{\csdef{#1}{\textsc{\lowercase{#1}}\xspace}}
\tsc{WGM}
\tsc{QE}
\tsc{EP}
\tsc{PMS}
\tsc{BEC}
\tsc{DE}
\begin{document}
\let\WriteBookmarks\relax
\def\floatpagepagefraction{1}
\def\textpagefraction{.001}
\shorttitle{Towards Segment Anything Model (SAM) for Medical Image Segmentation: A Survey}
\shortauthors{Ychi Zhang et~al.}

\title [mode = title]{Towards Segment Anything Model (SAM) for Medical Image Segmentation: A Survey}

\author[1]{Yichi Zhang}
\author[2]{Rushi Jiao}

\address[1]{School of Data Science, Fudan University, Shanghai, China}
\address[2]{School of Biomedical Engineering, Shanghai Jiao Tong University, Shanghai, China}

\begin{abstract}
Due to the flexibility of prompting, foundation models have become the dominant force in the domains of natural language processing and image generation. With the recent introduction of the Segment Anything Model (SAM), the prompt-driven paradigm has entered the realm of image segmentation, bringing with a range of previously unexplored capabilities. However, it remains unclear whether it can be applicable to medical image segmentation due to the significant differences between natural images and medical images.
In this work, we summarize recent efforts to extend the success of SAM to medical image segmentation tasks, including both empirical benchmarking and methodological adaptations, and discuss potential future directions for SAM in medical image segmentation.
Although directly applying SAM to medical image segmentation cannot obtain satisfying performance on multi-modal and multi-target medical datasets, many insights are drawn to guide future research to develop foundation models for medical image analysis. 
To facilitate future research, we maintain an active repository that contains up-to-date  paper list and open-source project summary at https://github.com/YichiZhang98/SAM4MIS.
\end{abstract}

%\includegraphics{figs/grabs.pdf}

%\item Research highlights item 1
%\item Research highlights item 2
%\item Research highlights item 3


\maketitle

\section{Introduction}

Medical image segmentation aims to distinguish specific anatomical structures including organs, lesions and tissues from medical images, which is a fundamental and essential step in numerous clinical applications, such as computer-aided diagnosis, treatment planning, and monitoring of disease progression \cite{MIA2017survey,AbdomenCT-1K}. Accurate segmentation can provide reliable volumetric and shape information of target structures, so as to assist in many further clinical applications like disease diagnosis, quantitative analysis and surgical planning.
Deep learning models have demonstrated great promise in the field of medical image segmentation due to their ability to learn intricate imaging features. 
However, existing deep models are often tailored for specific modalities and targets, which limits their capacity for further generalization to other tasks.

The emergence of large-scale foundation models \cite{wang2023large,liang2022foundations} has revolutionized artificial intelligence and sparked a new era due to their remarkable zero-shot and few-shot generalization abilities across a wide range of downstream tasks.
With the recent introduction of the Segment Anything Model (SAM) \cite{SAM-Meta} as an innovative foundational model for image segmentation, which has gained massive attention by its strong capabilities for generating accurate object masks in a fully automatic or interactive way. The prompt-driven paradigm has entered the realm of image segmentation, bringing with a range of previously unexplored capabilities.
However, as a very important branch of image segmentation, it remains unclear whether these foundation models can be applicable to medical image segmentation due to the significant differences between natural images and medical images.
To this end, a large number of extended works have been proposed by the community to further explore the usage of SAM to medical image segmentation tasks.

In this paper, we aim to summarize recent efforts to extend the success of SAM to medical image segmentation tasks.
Firstly, we briefly introduce the background of foundation models and the workflow of SAM. After that, we review and divide these works into two main categories. The first line of researches aim to evaluate the performance of SAM with different modes in different medical image segmentation tasks. Another line of researches focus on investigating methods to better adapt SAM to medical image segmentation tasks.
Finally, we conclude the survey and outline several existing challenges and potential future directions.
To keep up with the rapid increase of research community, we maintain a continuously updated paper list and open-source project summary to boost the research on this topic.

\begin{figure*}
	\includegraphics[width=18cm]{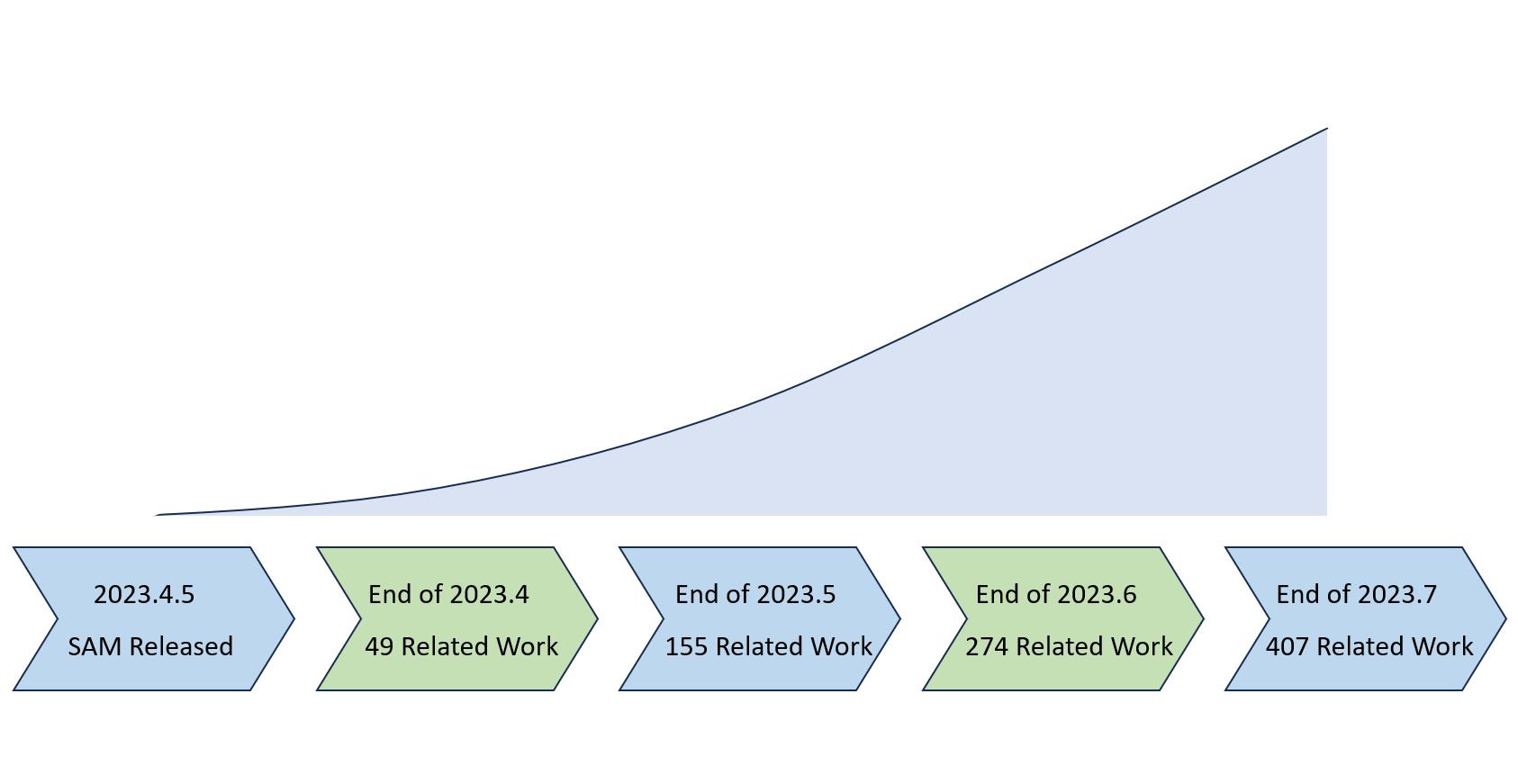}
	\caption{Statistics of papers related to Segment Anything Model (SAM) based on the citation count on Google Scholar in the fast-evolving field.}
	\label{trends}
\end{figure*}

\section{Background}

\subsection{Foundation Models}

Foundation models are a rapidly growing area of artificial intelligence research aimed at developing large-scale, general-purpose language and vision models. These models are often trained on massive amounts of data, which allows them to learn general representations and capabilities that can be transferred to different domains and applications.

One of the most widely known foundation models is the GPT (Generative Pre-trained Transformer) series \cite{GPT-3,GPT-4},  which demonstrated impressive capabilities and performance on a variety of natural language processing tasks such as sentence completion, question answering, and language translation.
These achievements have inspired researchers to develop large-scale foundational models to learn universial representations for computer vision tasks, which focus on capturing the cross-modal interactions between vision and language like understanding visual concepts and details \cite{CLIP}, generating natural language descriptions of image regions \cite{ALIGN} and generating images from textual descriptions \cite{DALL-E}. The success of these foundation models has spawned numerous derivative works and applications spanning different industries, which have become an essential component of many AI system architectures, and their continued development promises to drive further advances in language and vision tasks \cite{liu2023summary,awais2023foundational}.
Foundation models have also shown strong potential in solving a wide range of downstream tasks for medical image analysis and help to accelerate the development of accurate and robust models \cite{yi2023towards,zhang2023foundation,li2023artificial}.

\subsection{Segment Anything Model}

As the first promptable foundation model for image segmentation tasks, Segment Anything Model (SAM) is trained on the large-scale SA-1B dataset with an unprecedented number of images and annotations, which enables the model with strong zero-shot generalization.
SAM utilizes a transformer-based architecture \cite{attention-Nips17}, which has been shown to be highly effective in natural language processing \cite{GPT-3} and image recognition tasks \cite{ViT2020}.
Specifically, SAM uses vision transformer-based image encoder to extract image features and compute an image embedding, and prompt encoder to embed prompts and incorporate user interactions. Then extranted information from two encoders are combined to a lightweight mask decoder to generate segmentation results based on the image embedding, prompt embedding, and output token. 

\begin{figure*}
	\includegraphics[width=18cm]{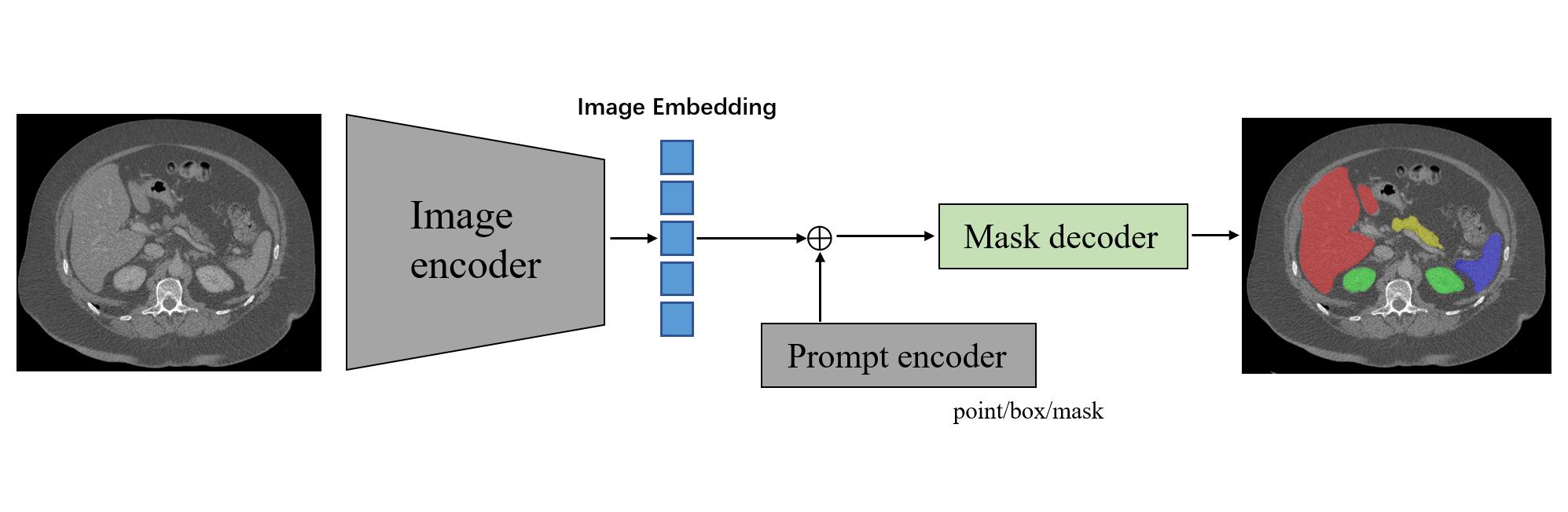}
	\caption{Overview of the architecture of Segment Anything Model (SAM).}
	\label{SAM}
\end{figure*}

\textbf{Image Encoder.} The vision transformer \cite{ViT2020} in the image encoder is pre-trained with masked auto-encoder \cite{MAE}, which is minimally adapted to process high-resolution (i.e. 1024$\times$1024) images. After image encoder, the obtained image embedding is 16$\times$ downscaled to 64$\times$64.

\textbf{Prompt Encoder.} For prompt encoders, two sets of prompts are considered, including sparse prompts (points, boxes, text) and dense prompts (masks).
SAM employs positional encoding \cite{FourierPE-Nips20} combined with learned embeddings to represent points and boxes. Specifically, points are encoded by two learnable tokens for specifying foreground and background, and the bounding box is encoded by the point encoding of its top-left corner and bottom-right corner. The free-form text is encoded by the pre-trained text-encoder from CLIP \cite{CLIP}. 
Dense mask prompts that has the same spatial resolution as the input image, are embedded using convolutions and summed element-wise with the image embedding.

\textbf{Mask Decoder.} The mask decoder is characterized by its lightweight design, which consists of two transformer layers with a dynamic mask prediction head and an Intersection-over-Union (IoU) score regression head. The mask prediction head is capable of producing three 4$\times$ downscaled masks, corresponding to the whole object, part, and subpart of the object, respectively.

During training, the output prediction is supervised with the linear combination of focal loss \cite{Lin2017FocalLF} and Dice loss \cite{Milletar2016VNetFC}. A data engine is built for label-efficient training. Specifically, professional annotators first label masks through interactive segmentation. Then, less prominent objects which are ignored in the predictions of SAM will be labeled manually. Finally, a fully automatic stage is conducted, in which confident and stable pseudo masks are selected as annotations.

\section{How SAM performs on Medical Image Segmentation? }

Despite SAM's impressive performance on natural images, it remains unclear whether it can be applicable to more challenging medical image segmentation due to structural complexity, low contrast and inter-observer variability.
To issue this problem, several studies have worked to answer the question that \textit{How SAM performs on Medical Image Segmentation?}

\textbf{Pathology Image Segmentation.} Deng \textit{et al.} \cite{SAM-pathology} evaluate SAM on tumor segmentation, non-tumor tissue segmentation and cell nuclei segmentation on whole slide imaging (WSI). By conducting several different scenarios including SAM with a single positive point prompt, with 20 point prompts with 10 positive and 10 negative points, and with all points/boxes on every single instance object, the results suggest that SAM achieves remarkable segmentation performance for large connected objects. However, SAM does
not achieve satisfying performance for dense instance object segmentation, even with 20 prompts (clicks/boxes) on each image. Possible reasons include significantly higher image resolution of WSI compared with the training image resolution of SAM, and multiple scales of different tissue types for digital pathology.

\textbf{Liver Tumor Segmentation from CECT.} Hu \textit{et al.} \cite{SAM-LiverTumor} conduct experiments on multi-phase liver tumor segmentation from contrast-enhanced computed tomography (CECT) volumes. From the results, it can be observed that the more prompt points used for segmentation, the better performance SAM can achieve. Compared with classic U-Net \cite{U-Net}, there is still a large gap when SAM with the few prompt points.

\textbf{Polyps Segmentation from Colonoscopy Images.} Zhou \textit{et al.} \cite{SAM-Polyps} evaluate the performance of SAM with unprompted setting in segmenting polyps from colonoscopy images on five benchmark datasets. From the experimental results, SAM demonstrates lower performance compared to state-of-the-art methods with an average Dice Similarity Coefficient (DSC) decrease ranging from 14.4\% to 36.9\%. When directly applying polyp segmentation task, SAM cannot achieve satisfactory performance for these unseen medical images due to blurred boundaries between a polyp and its surrounding mucosa.

\textbf{Brain MRI Segmentation.} Mohapatra \textit{et al.} \cite{SAM-BrainMR} compare SAM with Brain Extraction Tool (BET), which is a widely used and current gold standard technique for brain extraction and segmentation of magnetic resonance imaging. Experimental results show that SAM can obtain comparable or even better performance compared with BET, demonstrating significant additional potential to emerge as an efficient tool for brain extraction and segmentation applications.

\textbf{Abdominal CT Organ Segmentation.} Roy \textit{et al.} \cite{SAM-DKFZ-Abdomen} aim to evaluate the out-of-the-box zero-shot capabilities of SAM by conducting experiments on abdominal CT organ segmentation with scenarios including 1, 3 and 10 randomly selected points from the segmentation mask and bounding boxes of the segmentation masks with jitter of 0.01, 0.05, 0.1, 0.25 and 0.5 to simulate various degrees of user inaccuracy. Although SAM with point prompts underperform state-of-the-art performance with an average Dice Similarity Coefficient (DSC) decrease ranging from 20.3\% to 40.9\%, using box prompts can obtain highly competitive performance even with moderate (0.1) jitter.

\textbf{Endoscopic Surgical Instrument Segmentation.} Wang \textit{et al.} \cite{SAM-RS} evaluate the performance of SAM on two classical datasets in endoscopic surgical instrument segmentation. Experimental results suggest that SAM is deficient in segmenting the entire instrument with point-based prompts and unprompted settings. Specifically, SAM fails in predicting certain parts of the instrument when there are overlapping instruments or only with a point-based prompt. Besides, SAM also fails to identify instruments in complex surgical scenarios, such as blood, reflection, blur and shade.

\begin{figure*}
	\includegraphics[width=18cm]{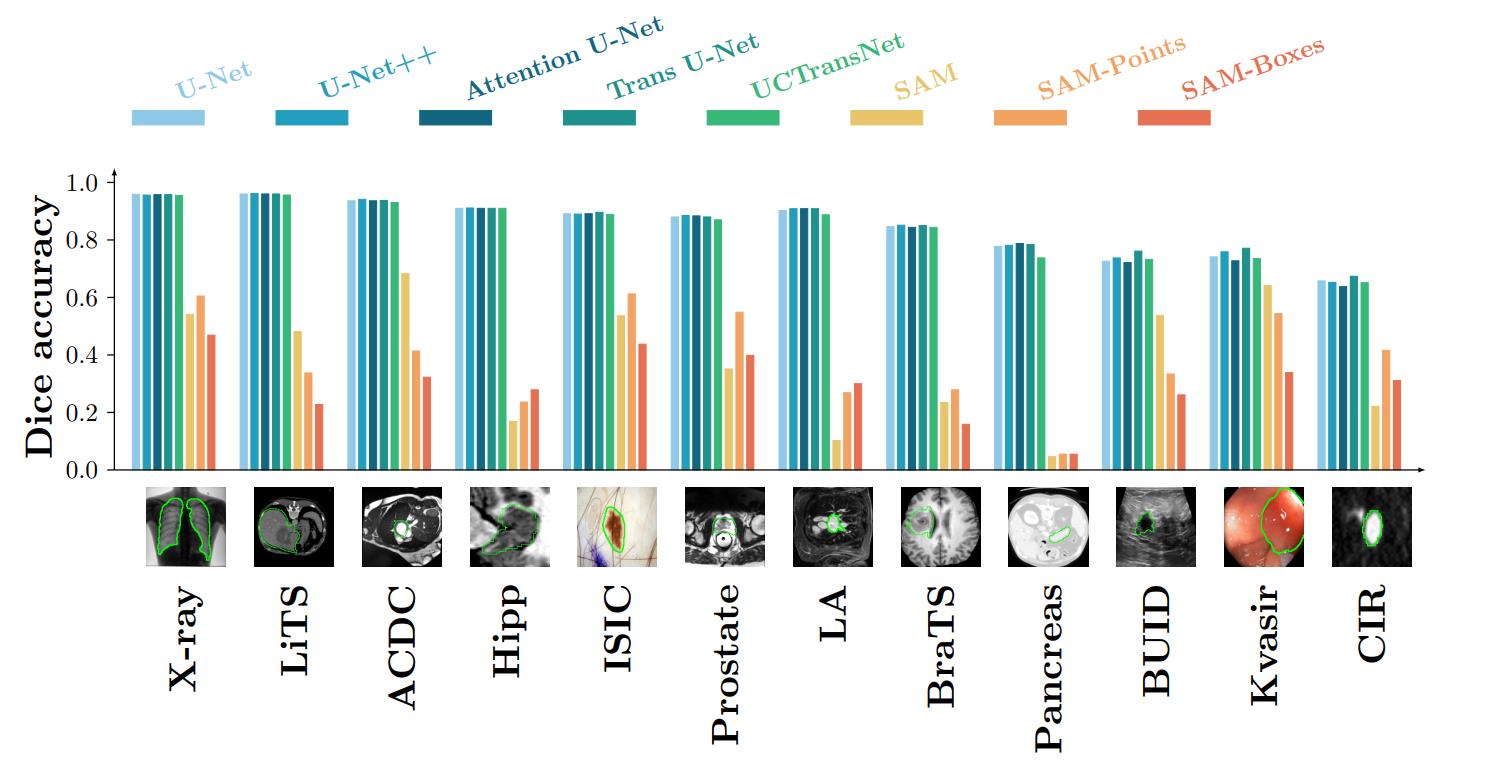}
	\caption{Visual comparison of Dice Similarity Coefficient between SAM and existing segmentation networks on 12 medical image segmentation datasets. \cite{SAM-Meds}}
	\label{Comparison}
\end{figure*}

\textbf{Other Multi-dataset Evaluations.} Instead of single-dataset evaluation, He \textit{et al.} \cite{SAM-Meds} conduct a large scale empirical study to evaluate the accuracy of SAM on 12 public medical image segmentation datasets covering different organs including brain, breast, chest, lung, skin, liver, bowel, pancreas, and prostate, different image modalities including 2D X-ray, histology, endoscopy, 3D MRI and CT, and different health conditions including normal, lesioned. Visual comparison of between SAM and existing segmentation networks can be found in Fig. \ref{Comparison}.
Similarly, Mazurowski \textit{et al.} \cite{SAM-Empirical} perform an extensive evaluation of SAM on a collection of 11 medical image segmentation datasets from different modalities including X-ray, ultrasound, MRI, CT and PET/CT, and various anatomies, by generating point prompts using a standard method to simulate interactive segmentation. The performance appears to be high for well-circumscribed objects with unambiguous prompts and poorer in other scenarios like tumor segmentation. Cheng \textit{et al.} \cite{SAM-MI} evaluate the performance of SAM on 12 public medical image datasets covering various organs and modalities under three prompt modes including auto-prompt mode, box-prompt mode and point-prompt mode. The experimental results show that the performance of SAM varies among different datasets, and the box-prompt mode without jitters is the best way to utilize the SAM model on medical image tasks compared with other prompt modes. Zhang \textit{et al.} \cite{SAM-RO} evaluate the performance of SAM in clinical radiotherapy for segmentation of prostate, lung, gastrointestinal, and head\&neck. The evaluation highlights SAM’s capacity to delineate large, distinct organs but remains challenging in segmenting smaller, intricate structures, especially with ambiguous prompts.
Overall, the segmentation results of SAM are generally lower than state-of-the-art methods, but there are still several datasets exceeding state-of-the-art methods.

Furthermore, Huang \textit{et al.} \cite{SAM-SZU} collect and sort 52 open-source datasets to built a large-scale medical segmentation dataset named COSMOS 553K with 16 modalities, 68 objects and 553K slice. Comprehensive experiments are conducted on different SAM testing strategies include everything modes, point-based and box-based prompt modes.
The experimental results validate that SAM performs better with manual prompts like points and boxes for object perception in medical images compared to everything mode. Specifically, by adding additional negative point prompts to positive point prompts which should theoretically improve the performance, it is observed that the addition of negative points will slightly decrease the performance on some tasks when background points are similar to foreground points. This finding indicates that point prompts should be selected carefully based on domain knowledge in order to achieve stable performance improvement.

\textbf{Summary.} 
In this section, we review recent studies to benchmark SAM on different medical image segmentation tasks with comparison to existing domain-specific segmentation methods.
Generally, SAM requires substantial human information to obtain overall moderate segmentation performance using only few point or bounding box prompts.
Overall, these evaluation results on different datasets show that SAM has limited generalization ability when directly applying to medical image segmentation, which varies significantly across different datasets and tasks.
SAM shows remarkable performance comparable to state-of-the-art methods in some specific objects and modalities. However, SAM is imperfect or even totally fails in other more challenging situations when the segmentation targets have weak boundaries with low-contrast and smaller and irregular shape, which is in-line with other investigations \cite{SAM-ConcealedScenes,SAM-Realworld}. 
For most situations, the subpar segmentation performance of SAM is not sufficient and satisfying for further applications, especially for medical image segmentation tasks where extremely high accuracy are demanded.
Since SAM is pre-trained on SA-1B dataset consists of natural images where the objects usually have strong edge information, which is significantly different from medical images. Therefore, directly applying SAM to these unseen and challenging medical image segmentation tasks may have limited performance.

\section{How to better adapt SAM to Medical Image Segmentation?}

Several studies have shown that SAM could fail on medical image segmentation tasks especially when the segmentation targets have weak boundaries. To better adapt SAM to medical image segmentation tasks, another line of researches have worked on these following topics.

\textbf{Fine-tuning SAM on Medical Datasets.} 
Due to the unsatisfying performance of directly applying SAM to medical image segmentation, several researches investigate how to fine-tune SAM on medical images to improve its reliability.
Since updating all the parameters of SAM is time-consuming, computationally intensive and difficult to deploy, most researches focus on fine-tuning a small fraction of parameters of SAM, such as one of the sub-parts of SAM (image encoder, prompt encoder and mask decoder) or the combinations of them. Hu \textit{et al.} \cite{skinSAM} evaluate the performance of directly applying and fine-tuning SAM except the image encoder to skin cancer segmentation task. By fine-tuning SAM on target datasets, the model can obtain significant performance gains from 81.25\% to 88.79\% in Dice. Li \textit{et al.} \cite{PolypSAM} propose to transfer SAM to polyp segmentation task with fine-tuning. Experimental results demonstrate excellent performance on five public datasets when compared to existing methods. 
Ma \textit{et al.} \cite{MedSAM} introduce MedSAM for universal image segmentation by curating a diverse and comprehensive medical image dataset containing over 200,000 masks with 11 modalities and develop fine-tuning approach to adapt SAM to medical image segmentation. The proposed MedSAM further improve the performance of SAM with an average Dice Similarity Coefficient (DSC) of 22.51\% on 21 3D segmentation tasks, and of 17.61\% on 9 2D segmentation tasks, demonstrating the effectiveness of fine-tuning the mask encoder on medical images. However, the overall performance is still behind specialist models for medical image segmentation.
Wu \textit{et al.} \cite{Med-SAM-Adapter} introduce Medical SAM Adapter (MSA) to fine-tuning pre-trained SAM with a parameter-efficient fine-tuning paradigm using Adaption modules \cite{Lora}, since it enables efficient learning with faster updates and avoids catastrophic forgetting. Specifically, in SAM encoder, two adapters for each ViT block and the embedding is scaled, while in SAM decoder, three adapters for each ViT block.
Besides, other than classic 2D image segmentation, to utilize spatial correlation between slices for 3D medical image segmentation, MSA splits the attention operation into the space branch and the depth branch to further utilize the depth correlations, then the results from the depth branch are transposed back to original shape and added to the results of the space branch.
Comprehensive experiments on 19 medical image segmentation tasks across 5 different modalities including CT, MRI, ultrasound, fundus images and dermoscopic images demonstrate that by fine-tuning, MSA can obtain comparable performance compared with state-of-the-art methods.
Similarly, Zhang \textit{et al.} \cite{SAMed} apply low-rank-based (LoRA) \cite{Lora} fine-tuning strategy to SAM image encoder together with the prompt encoder and mask decoder on labeled medical image segmentation datasets. By fine-tuning on a multi-organ segmentation dataset, SAM can achieve highly competitive segmentation performance compared with state-of-the-art methods.
Gong \textit{et al.} \cite{3DSAM-adapter} propose to utilize a holistic modification scheme to transfer SAM from 2D to 3D for medical image segmentation while reusing most of pre-trained parameters. SPecifically, the fine-tuning process is conducted in a parameter-efficient manner, wherein most of the pre-trained parameters are kept frozen, while only a few lightweight spatial adapters are introduced.
Zhang \textit{et al.} \cite{SAM-Path} introduce a fine-tuning approach using trainable class prompts to identify classes using SAM for pathology segmentation. Specifically, an integration of a pathology encoder to incorporate more domain-specific knowledge.
Chai \textit{et al.} \cite{chai2023ladder} combine an additional CNN as a complementary encoder along with the standard SAM architecture and only focus on fine-tuning the additional CNN and SAM decoder to reduce the resource utilization and training time of fine-tuning. 
For multi-modal segmentation tasks, Shi \textit{et al.} propose a cross-modality attention adapter based on multi-modal fusion for fine-tuning SAM for glioma segmentation from multi-modal MRI images.

As a simple and directive approach, these tuning-based methods demonstrate the effectiveness of fine-tuning SAM on domain-specific medical datasets to achieve better segmentation performance.

\begin{figure}
	\includegraphics[width=8cm]{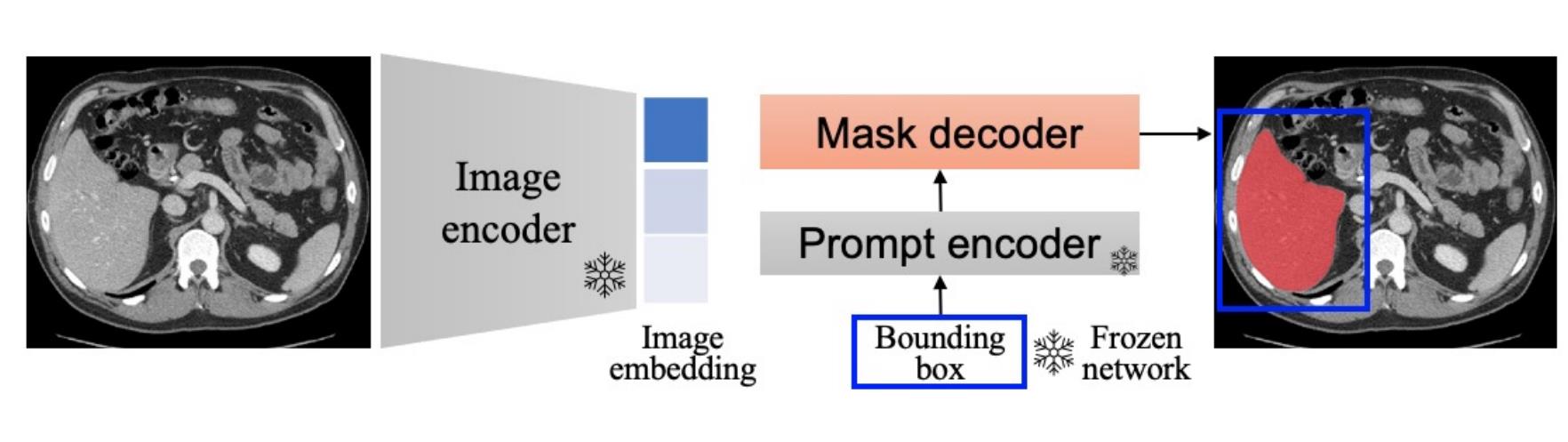}
	\caption{MedSAM: Fine-tuning Segment Anything Model for medical image segmentation by freezing the image encoder and prompt encoder and only fine-tuning the mask decoder. \cite{MedSAM}.}
	\label{MedSAM}
\end{figure}

\begin{figure}
	\includegraphics[width=8cm]{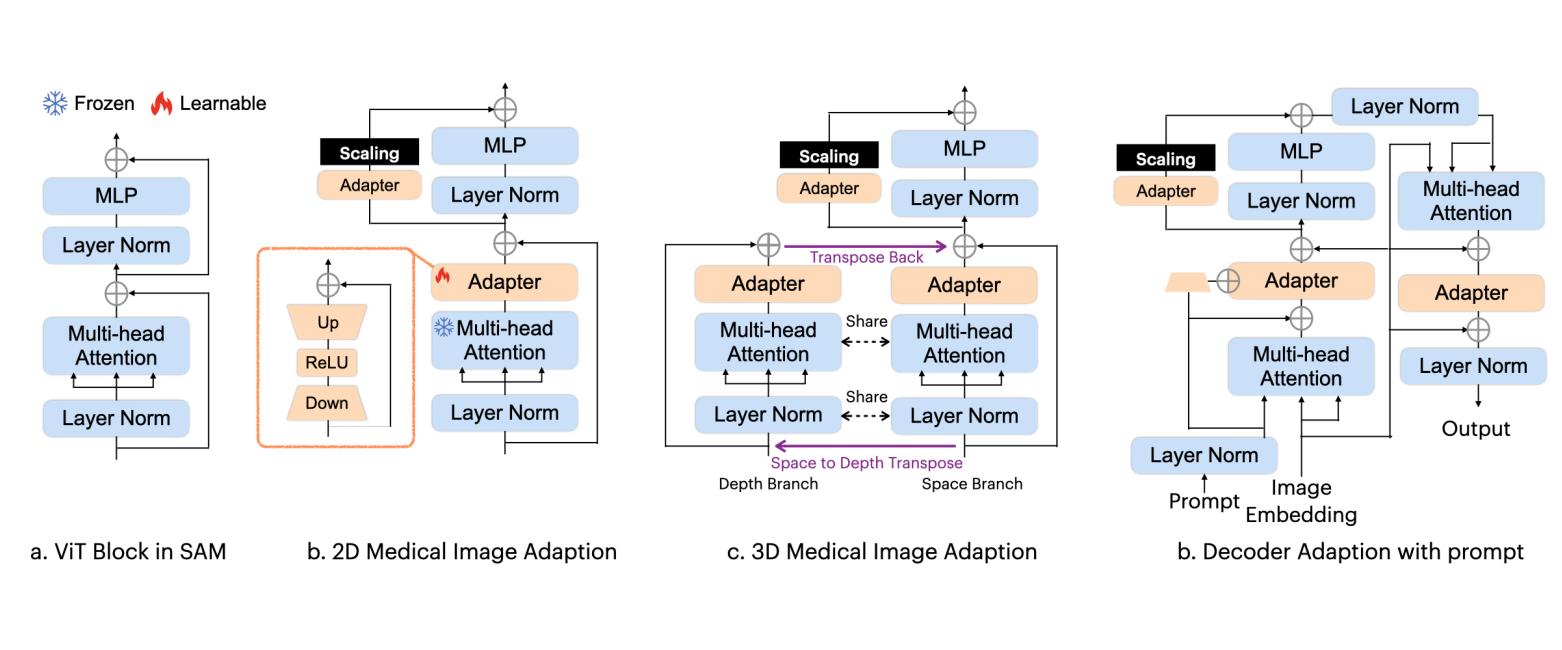}
	\caption{The architecture of Medical SAM Adapter for 2D and 3D medical images. \cite{Med-SAM-Adapter}.}
	\label{MSA}
\end{figure}

\textbf{Extending the Usability of SAM to Medical Images.}
Different from classic natural images, medical images are mostly in specific formats such as NII and DICOM. To ease the the usage and assist in the utilization of SAM to medical images, Liu \textit{et al.} \cite{SAM-3DSlicer} extend SAM into commonly used medical image viewers with 3D Slicer \cite{3Dslicer}, which enables researchers to conduct segmentation on medical images with only a latency of 0.6 seconds. The segmentation process through prompt can be automatically applied to the next slice when a slice’s segmentation is complete.

While the performance of SAM is conditioned by input prompts, it may be desirable to have a fully automatic solution.
Shaharabany \textit{et al.} \cite{AutoSAM} propose AutoSAM to involve the training of an auxiliary prompt encoder to generate a surrogate prompt. 
Beyond the classic prompt types of SAM including bounding box, points or masks, AutoSAM utilizes the image itself as its input. During training, the model propagates gradients to the prompt encoder with binary cross-entropy loss and Dice loss. 
With the guidance of auxiliary trained network, SAM is truned into a fully automatic manner since no prompt is required, and obtain state-of-the-art results on multiple medical benchmarks without fine-tuning.
Cui \textit{et al.} \cite{All-in-SAM} introduce a pipeline named all-in-SAM to utilizes SAM without requiring manual prompts. Specifically, the framework leverages weak annotations and pre-trained SAM for fine-tuning to minimize annotation costs, therefore enhancing the application of SAM through label-efficient fine-tuning. Lei \textit{et al.} \cite{MedLSAM} propose MedLSAM to apply a localization process by identifying six extreme points in three directions of 3D images of any region of interest. After that, the generated bounding box is utilized by SAM to carry out precise segmentation of the target anatomy for sutomatic segmentation.

\textbf{Enhancing the Robustness against Different Prompts.}
Although fine-tuning SAM on medical datasets can achieve performance improvement, it still requires the use of manually given boxes or points, making it difficult to achieve fully automatic medical image segmentation. While for prompt mode, the final segmentation result is highly dependent on the prompt, and the model still tends to be more sensitive to wrong prompts.
To issue this challenge, Gao \textit{et al.} \cite{DeSAM} propose Decoupling Segment Anything Model (DeSAM) by decoupling the mask decoder of SAM into two subtasks: prompt-relevant IoU module (PRIM) to generate mask embeddings based on given prompt, and prompt-invariant mask module (PIMM) to fused the image embeddings from the image encoder with the mask embeddings from PRIM to generate the mask. 
DeSAM can minimize the performance degradation caused by wrong prompts while avoiding training image encoder which requires higher GPU cost. Extensive experiments demonstrate that DeSAM improves the robustness of fully automated segmentation in dealing with distribution variations across different sites.
Deng \textit{et al.} \cite{SAM-U} propose to enhance SAM by employing multiple box prompts to establish pixel-level reliability through uncertainty estimation. By generating different predictions using different multi-box prompts and estimating the distribution of SAM predictions using Monte Carlo simulation with prior distribution parameters, the model can estimate the variations by aleatoric uncertainty and generate an uncertainty maps to highlight challenging areas for segmentation, which offers valuable guidance for potential segmentation errors and support further clinical analysis.

\begin{figure}
	\includegraphics[width=8cm]{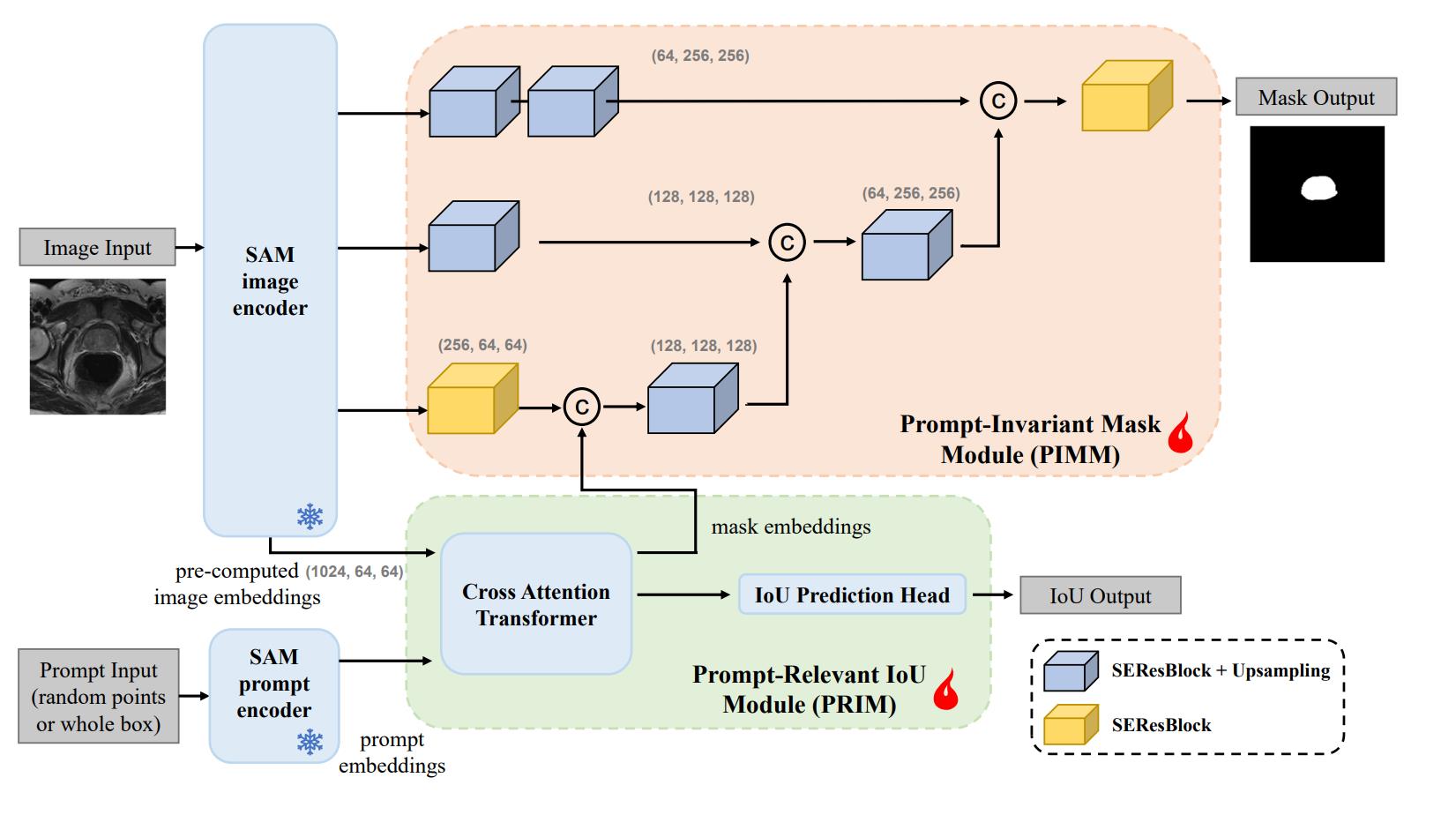}
	\caption{DeSAM: Decoupling Segment Anything Model with Prompt-Relevant IoU Module (PRIM) to generate mask embeddings based on given prompt and Prompt-Pnvariant Mask Module (PIMM) to fused the image embeddings with the mask embeddings to generate the mask. \cite{DeSAM}.}
	\label{DeSAM}
\end{figure}

\begin{figure}
	\includegraphics[width=8cm]{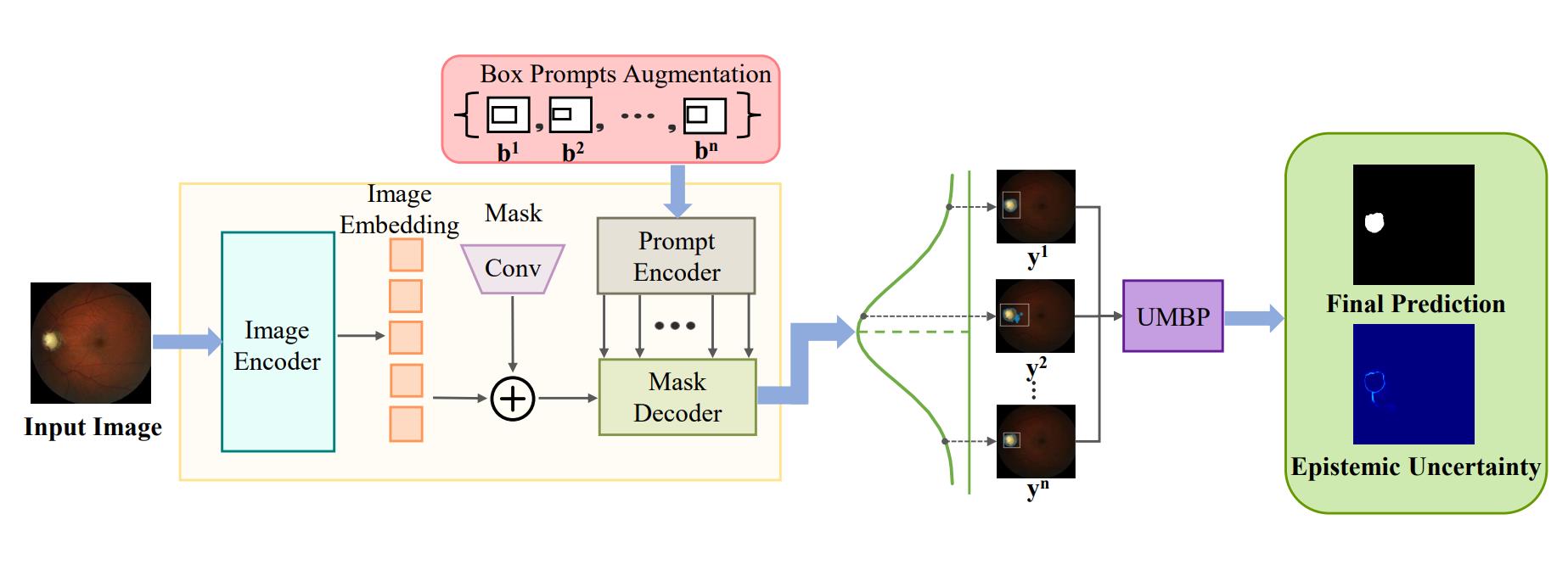}
	\caption{The architecture of SAM-U with multi-box prompts and uncertainty estimation \cite{SAM-U}.}
	\label{SAM-U}
\end{figure}

\textbf{Input Augmentation with SAM.}
Due to SAM's imperfect performance when directly applying to medical image segmentation tasks that require domain-specific knowledge, instead of directly applying SAM for segmentation, Zhang \textit{et al.} \cite{IA-SAM} aim to directly utilize the segmentation masks generated by SAM to augment the raw input medical images. The input augmentation is performed by a fusion function SAMAug to generate segmentation prior map and boundary prior map. Specifically, SAM uses a grid prompt to generate segmentation masks with all plausible locations in the image. Then the masks are drawn to form a segmentation prior map with the corresponding stability scores of masks generated by SAM. In addition to the segmentation prior map, the boundary prior map is further generated according to the masks
provided by SAM. With the prior maps generated, the input image is augmented by adding the prior maps to the raw image with multiple channels.
Experiments on two datasets show that using the same U-Net for segmentation, adapting SAMAug for input augmentation can improve the Aggregated Jaccard Index (AJI) from 58.36\% to 64.30\% on cell segmentation task, and Object Dice from 86.35\% to 87.44\% on gland segmentation task. 
These result proves that although SAM may not generate high-quality segmentation for medical images, these generated masks and features are still useful to boost the segmentation models.

\section{Discussion and Conclusion}

In this paper, we summarize recent efforts of applying SAM to medical image segmentation tasks. 
When directly applying SAM to medical image segmentation, the segmentation performance varies significantly across different datasets and tasks, which represents that SAM cannot stably and accurately implement zero-shot segmentation on multi-modal and multi-target medical datasets. 
Different attributes of the segmentation targets and imaging modalities may affect SAM’s segmentation ability. Particularly, SAM may output poor results or even totally fails for objects with irregular shapes, weak boundaries, small sizes, or low contrast,
For most situations, the subpar segmentation performance of SAM is not sufficient and satisfying especially for medical image segmentation where extremely high accuracy are demanded.
Besides, due to the significant differences between natural images and medical images, sever researches \cite{MedSAM,Med-SAM-Adapter} prove that suitable fine-tuning strategy on medical images can improve the unsatisfactory segmentation results to some extent, so as to achieve competitive performance compared with domain-specific models.

In conclusion, the success of SAM demonstrates the feasibility of building segmentation foundation models.
Although SAM's performance is not stable compared with domain-specific models at present, we believe that it has the strong potential to serve as an efficient and powerful tool to further build segmentation models so as to assist in clinical applications.
We outline some of existing challenges and potential future directions as follows.

\textbf{Buiding Large-Scale Medical Datasets.}
Although fine-tuning SAM on medical datasets can improve the segmentation results, it still has performance limitations in medical image segmentation tasks due to the differences between natural images and medical images.
Therefore, it is of great importance to build large-scale medical datasets for developing general segmentation models specifically in the medical domain.
For comparison, the SA-1B dataset used for training SAM consists of 11M images with 1.1B segmentation masks \cite{SAM-Meta}, while for medical imaging, it is labour-intensive and expertise-demanding to collect large-scale labeled datasets since only experts can provide reliable and accurate annotations. Several recent studies have focus on creating large-scale medical datasets \cite{SAM-SZU,qu2023annotating}, which may contribute to the future development of foundation models for medical image segmentation \cite{moor2023foundation,willemink2022toward}.

\textbf{Integrating Clinical Domain Knowledge.}
SAM could be improved by integrating clinical information, which is useful other than image information. For example, the relationship between tumor location, size, and the expected effect could be used to assist in the segmentation procedure.
Combining this with SAM’s powerful image segmentation capabilities could be both accurate and clinically relevant.

\textbf{Adapt SAM from 2D to 3D Medical Images.}
A major challenge for adapting SAM to medical image segmentation is the difference in the image dimension. Unlike classic 2D natural images, many medical scans are 3D volumes such as MRI and CT. For these 3D medical images, physicians need to utilize the correlation between adjacent slices to make decisions.
Although SAM can be applied to 3D medical images by segmenting each slice of the volume to generate the final segmentation result, the inter-slice information between adjacent slices is neglected. Many previous studies have shown the importance of inter-slice correlation for identification of some objects so as to ensure accurate segmentation \cite{2-5D}.
Several recent studies have also focused on transferring SAM from 2D to 3D for medical image segmentation to support volumetric inputs to issue this challenge \cite{Med-SAM-Adapter,3DSAM-adapter}.

\begin{figure}
	\includegraphics[width=8cm]{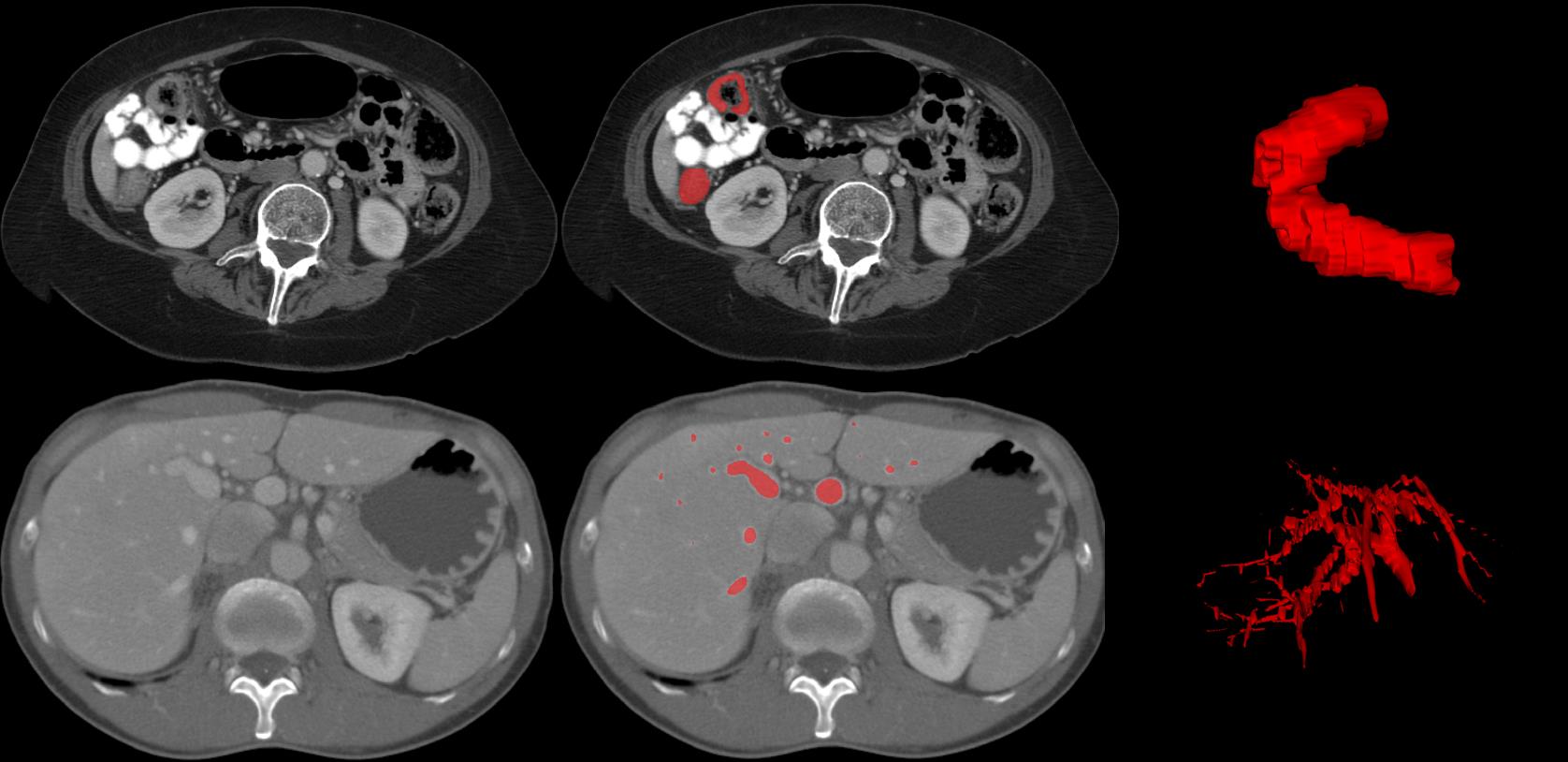}
	\caption{Examples of colon segmentation and hepatic Vessel segmentation from CT volumes, where volumetric information between 2D slices is important to ensure accurate segmentation.}
	\label{2D-3D}
\end{figure}

\textbf{Reduce the Annotation Cost for Medical Image Segmentation.}
The heavy annotation cost of medical images is one of the major challenges of developing segmentation models, since it typically requires the domain-specific knowledge of experts to provide reliable and accurate annotations \cite{SemiSurvey,MIA-Imperfect}.
Although the segmentation results generated by SAM is not always perfect, these segmentation masks can still be considered to reduce the annotation cost. Instead of labeling targets from scratch, experts can utilize SAM to implement a coarse segmentation and then revise the segmentation manually, so as to achieve fast interactive segmentation. 
This direction is also explored by several recent studies. For example, Wang \textit{et al.} propose $SAM^{Med}$ framework for medical image annotation to leverage the capabilities of SAM, which consists of two submodules for automatic generation of annotations with $SAM^{auto}$ and assisting users for efficient annotating medical images with $SAM^{assist}$. Shen \textit{et al.} \cite{SAM--Med} focus on utilizing SAM's zero-shot capabilities for interactive medical image segmentation with an innovative reinforcement learning-based framework named temporally-extended prompts optimization (TEPO) by adaptively providing suitable prompt forms for human experts.
Huang \textit{et al.} \cite{huang2023push} propose a label corruption framework to push the boundary of SAM-based segmentation for pseodu-label correction by utilizing a novel noise detection module to distinguish between noisy labels from clean labels with uncertainty-based self-correction.

\begin{figure}
	\includegraphics[width=8cm]{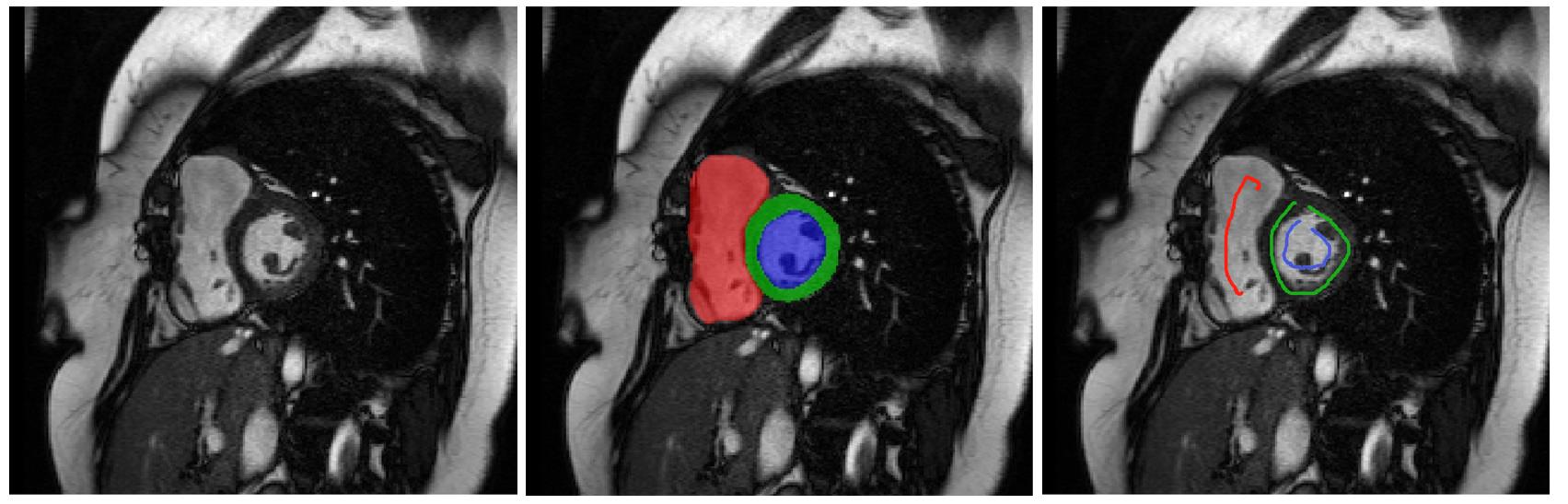}
	\caption{Visual comparison of fully and scribble annotations for cardiac MRI segmentation, where box prompts may not be suitable.}
	\label{Scribble}
\end{figure}

\textbf{Beyond Point and Box Prompts.}
Several empirical studies \cite{SAM-MI,SAM-DKFZ-Abdomen} illustrate that using box prompts can achieve better results compared with point prompts because relatively more accurate location information can be obtained. However, if there are multiple similar instances around the segmentation target, large bounding box may confuse the model and lead to incorrect segmentation results.
In addition to the point and box prompts, scribble prompt is another widely used interaction for medical image segmentation \cite{Scribbleseg-1,Scribbleseg-2}, which is useful and efficient for some non-compact targets with irregular shape and may further improve the performance when incorporated to SAM as illustrated in Fig.\ref{Scribble}. 

\textbf{Integration into Clinical Workflow.}
In addition to existing methodological adaption of SAM for medical image segmentation tasks, researchers could also focus on integrating SAM  to downstream tasks based on image segmentation to assist in the clinical workflow.
For example, Wang \textit{et al.} \cite{GazeSAM} investigate the potential integration of SAM and eye-tracking technology to design a collaborative human-computer interaction system and enable radiologists to acquire segmentation masks by simply looking at the region of interest. 
Ning \textit{et al.} \cite{SAM-UIG} discuss the potential contribution of SAM to enable universal intelligent ultrasound image guidance.
SAM can also provide preliminary segmentation and select out complex cases that require more in-depth scrutiny by the clinicians, so as to reduce the burden on clinical experts. Besides, SAM could aid in minimizing inter-observer variability, which is a prevalent issue in manual contouring \cite{lappas2022interobserver}.

To embrace SAM in the era of big data and foundation models, there is more we can do.
We hope this work can provide the community with some insights into the future development of foundation models for medical image segmentation.

\section*{Declaration of Competing Interest}
The authors have no conflict of interest to disclose.

\bibliographystyle{IEEEtran}
\bibliography{ref}

\begin{thebibliography}{10}
\providecommand{\url}[1]{#1}
\csname url@samestyle\endcsname
\providecommand{\newblock}{\relax}
\providecommand{\bibinfo}[2]{#2}
\providecommand{\BIBentrySTDinterwordspacing}{\spaceskip=0pt\relax}
\providecommand{\BIBentryALTinterwordstretchfactor}{4}
\providecommand{\BIBentryALTinterwordspacing}{\spaceskip=\fontdimen2\font plus
\BIBentryALTinterwordstretchfactor\fontdimen3\font minus
  \fontdimen4\font\relax}
\providecommand{\BIBforeignlanguage}[2]{{%
\expandafter\ifx\csname l@#1\endcsname\relax
\typeout{** WARNING: IEEEtran.bst: No hyphenation pattern has been}%
\typeout{** loaded for the language `#1'. Using the pattern for}%
\typeout{** the default language instead.}%
\else
\language=\csname l@#1\endcsname
\fi
#2}}
\providecommand{\BIBdecl}{\relax}
\BIBdecl

\bibitem{MIA2017survey}
G.~Litjens, T.~Kooi, B.~E. Bejnordi, A.~A.~A. Setio, F.~Ciompi, M.~Ghafoorian,
  J.~A. Van Der~Laak, B.~Van~Ginneken, and C.~I. S{\'a}nchez, ``A survey on
  deep learning in medical image analysis,'' \emph{Medical image analysis},
  vol.~42, pp. 60--88, 2017.

\bibitem{AbdomenCT-1K}
J.~Ma, Y.~Zhang, S.~Gu, C.~Zhu, C.~Ge, Y.~Zhang, X.~An, C.~Wang, Q.~Wang,
  X.~Liu, S.~Cao, Q.~Zhang, S.~Liu, Y.~Wang, Y.~Li, J.~He, and X.~Yang,
  ``Abdomenct-1k: Is abdominal organ segmentation a solved problem?''
  \emph{IEEE Transactions on Pattern Analysis and Machine Intelligence},
  vol.~44, no.~10, pp. 6695--6714, 2022.

\bibitem{wang2023large}
X.~Wang, G.~Chen, G.~Qian, P.~Gao, X.-Y. Wei, Y.~Wang, Y.~Tian, and W.~Gao,
  ``Large-scale multi-modal pre-trained models: A comprehensive survey,''
  \emph{Machine Intelligence Research}, pp. 1--36, 2023.

\bibitem{liang2022foundations}
P.~P. Liang, A.~Zadeh, and L.-P. Morency, ``Foundations and recent trends in
  multimodal machine learning: Principles, challenges, and open questions,''
  \emph{arXiv preprint arXiv:2209.03430}, 2022.

\bibitem{SAM-Meta}
A.~Kirillov, E.~Mintun, N.~Ravi, H.~Mao, C.~Rolland, L.~Gustafson, T.~Xiao,
  S.~Whitehead, A.~C. Berg, W.-Y. Lo \emph{et~al.}, ``Segment anything,''
  \emph{arXiv preprint arXiv:2304.02643}, 2023.

\bibitem{GPT-3}
T.~Brown, B.~Mann, N.~Ryder, M.~Subbiah, J.~D. Kaplan, P.~Dhariwal,
  A.~Neelakantan, P.~Shyam, G.~Sastry, A.~Askell \emph{et~al.}, ``Language
  models are few-shot learners,'' \emph{Advances in neural information
  processing systems}, vol.~33, pp. 1877--1901, 2020.

\bibitem{GPT-4}
OpenAI, ``Gpt-4 technical report,'' \emph{arXiv preprint arXiv:2303.08774},
  2023.

\bibitem{CLIP}
A.~Radford, J.~W. Kim, C.~Hallacy, A.~Ramesh, G.~Goh, S.~Agarwal, G.~Sastry,
  A.~Askell, P.~Mishkin, J.~Clark \emph{et~al.}, ``Learning transferable visual
  models from natural language supervision,'' in \emph{International conference
  on machine learning}.\hskip 1em plus 0.5em minus 0.4em\relax PMLR, 2021, pp.
  8748--8763.

\bibitem{ALIGN}
C.~Jia, Y.~Yang, Y.~Xia, Y.-T. Chen, Z.~Parekh, H.~Pham, Q.~Le, Y.-H. Sung,
  Z.~Li, and T.~Duerig, ``Scaling up visual and vision-language representation
  learning with noisy text supervision,'' in \emph{International Conference on
  Machine Learning}.\hskip 1em plus 0.5em minus 0.4em\relax PMLR, 2021, pp.
  4904--4916.

\bibitem{DALL-E}
A.~Ramesh, M.~Pavlov, G.~Goh, S.~Gray, C.~Voss, A.~Radford, M.~Chen, and
  I.~Sutskever, ``Zero-shot text-to-image generation,'' in \emph{International
  Conference on Machine Learning}.\hskip 1em plus 0.5em minus 0.4em\relax PMLR,
  2021, pp. 8821--8831.

\bibitem{liu2023summary}
Y.~Liu, T.~Han, S.~Ma, J.~Zhang, Y.~Yang, J.~Tian, H.~He, A.~Li, M.~He, Z.~Liu
  \emph{et~al.}, ``Summary of chatgpt/gpt-4 research and perspective towards
  the future of large language models,'' \emph{arXiv preprint
  arXiv:2304.01852}, 2023.

\bibitem{awais2023foundational}
M.~Awais, M.~Naseer, S.~Khan, R.~M. Anwer, H.~Cholakkal, M.~Shah, M.-H. Yang,
  and F.~S. Khan, ``Foundational models defining a new era in vision: A survey
  and outlook,'' \emph{arXiv preprint arXiv:2307.13721}, 2023.

\bibitem{yi2023towards}
H.~Yi, Z.~Qin, Q.~Lao, W.~Xu, Z.~Jiang, D.~Wang, S.~Zhang, and K.~Li, ``Towards
  general purpose medical ai: Continual learning medical foundation model,''
  \emph{arXiv preprint arXiv:2303.06580}, 2023.

\bibitem{zhang2023foundation}
S.~Zhang and D.~N. Metaxas, ``On the challenges and perspectives of foundation
  models for medical image analysis,'' \emph{arXiv preprint arXiv:2306.05705},
  2023.

\bibitem{li2023artificial}
X.~Li, L.~Zhang, Z.~Wu, Z.~Liu, L.~Zhao, Y.~Yuan, J.~Liu, G.~Li, D.~Zhu, P.~Yan
  \emph{et~al.}, ``Artificial general intelligence for medical imaging,''
  \emph{arXiv preprint arXiv:2306.05480}, 2023.

\bibitem{attention-Nips17}
A.~Vaswani, N.~Shazeer, N.~Parmar, J.~Uszkoreit, L.~Jones, A.~N. Gomez,
  {\L}.~Kaiser, and I.~Polosukhin, ``Attention is all you need,''
  \emph{Advances in neural Information Processing Systems}, vol.~30, 2017.

\bibitem{ViT2020}
A.~Dosovitskiy, L.~Beyer, A.~Kolesnikov, D.~Weissenborn, X.~Zhai,
  T.~Unterthiner, M.~Dehghani, M.~Minderer, G.~Heigold, S.~Gelly \emph{et~al.},
  ``An image is worth 16x16 words: Transformers for image recognition at
  scale,'' in \emph{International Conference on Learning Representations},
  2020.

\bibitem{MAE}
K.~He, X.~Chen, S.~Xie, Y.~Li, P.~Doll{\'a}r, and R.~Girshick, ``Masked
  autoencoders are scalable vision learners,'' in \emph{Proceedings of the
  IEEE/CVF Conference on Computer Vision and Pattern Recognition}, 2022, pp.
  16\,000--16\,009.

\bibitem{FourierPE-Nips20}
M.~Tancik, P.~Srinivasan, B.~Mildenhall, S.~Fridovich-Keil, N.~Raghavan,
  U.~Singhal, R.~Ramamoorthi, J.~Barron, and R.~Ng, ``Fourier features let
  networks learn high frequency functions in low dimensional domains,''
  \emph{Advances in Neural Information Processing Systems}, vol.~33, pp.
  7537--7547, 2020.

\bibitem{Lin2017FocalLF}
T.-Y. Lin, P.~Goyal, R.~B. Girshick, K.~He, and P.~Doll{\'a}r, ``Focal loss for
  dense object detection,'' \emph{2017 IEEE International Conference on
  Computer Vision (ICCV)}, pp. 2999--3007, 2017.

\bibitem{Milletar2016VNetFC}
F.~Milletar{\`i}, N.~Navab, and S.-A. Ahmadi, ``V-net: Fully convolutional
  neural networks for volumetric medical image segmentation,'' \emph{2016
  Fourth International Conference on 3D Vision (3DV)}, pp. 565--571, 2016.

\bibitem{SAM-pathology}
R.~Deng, C.~Cui, Q.~Liu, T.~Yao, L.~W. Remedios, S.~Bao, B.~A. Landman, L.~E.
  Wheless, L.~A. Coburn, K.~T. Wilson \emph{et~al.}, ``Segment anything model
  (sam) for digital pathology: Assess zero-shot segmentation on whole slide
  imaging,'' \emph{arXiv preprint arXiv:2304.04155}, 2023.

\bibitem{SAM-LiverTumor}
C.~Hu and X.~Li, ``When sam meets medical images: An investigation of segment
  anything model (sam) on multi-phase liver tumor segmentation,'' \emph{arXiv
  preprint arXiv:2304.08506}, 2023.

\bibitem{U-Net}
O.~Ronneberger, P.~Fischer, and T.~Brox, ``U-net: Convolutional networks for
  biomedical image segmentation,'' in \emph{International Conference on Medical
  image computing and computer-assisted intervention}.\hskip 1em plus 0.5em
  minus 0.4em\relax Springer, 2015, pp. 234--241.

\bibitem{SAM-Polyps}
T.~Zhou, Y.~Zhang, Y.~Zhou, Y.~Wu, and C.~Gong, ``Can sam segment polyps?''
  \emph{arXiv preprint arXiv:2304.07583}, 2023.

\bibitem{SAM-BrainMR}
S.~Mohapatra, A.~Gosai, and G.~Schlaug, ``Sam vs bet: A comparative study for
  brain extraction and segmentation of magnetic resonance images using deep
  learning,'' \emph{arXiv preprint arXiv:2304.04738}, 2023.

\bibitem{SAM-DKFZ-Abdomen}
S.~Roy, T.~Wald, G.~Koehler, M.~R. Rokuss, N.~Disch, J.~Holzschuh, D.~Zimmerer,
  and K.~H. Maier-Hein, ``Sam. md: Zero-shot medical image segmentation
  capabilities of the segment anything model,'' \emph{arXiv preprint
  arXiv:2304.05396}, 2023.

\bibitem{SAM-RS}
A.-C. Wang, M.~Islam, M.~Xu, Y.~Zhang, and H.~Ren, ``Sam meets robotic surgery:
  An empirical study in robustness perspective,'' \emph{arXiv preprint
  arXiv:2304.14674}, 2023.

\bibitem{SAM-Meds}
S.~He, R.~Bao, J.~Li, P.~E. Grant, and Y.~Ou, ``Accuracy of segment-anything
  model (sam) in medical image segmentation tasks,'' \emph{arXiv preprint
  arXiv:2304.09324}, 2023.

\bibitem{SAM-Empirical}
A.~M. Maciej, D.~Haoyu, G.~Hanxue, Y.~Jichen, K.~Nicholas, and Z.~Yixin,
  ``Segment anything model for medical image analysis: an experimental study,''
  \emph{arXiv preprint arXiv:2304.10517}, 2023.

\bibitem{SAM-MI}
D.~Cheng, Z.~Qin, Z.~Jiang, S.~Zhang, Q.~Lao, and K.~Li, ``Sam on medical
  images: A comprehensive study on three prompt modes,'' \emph{arXiv preprint
  arXiv:2305.00035}, 2023.

\bibitem{SAM-RO}
L.~Zhang, Z.~Liu, L.~Zhang, Z.~Wu, X.~Yu, J.~Holmes, H.~Feng, H.~Dai, X.~Li,
  Q.~Li \emph{et~al.}, ``Segment anything model (sam) for radiation oncology,''
  \emph{arXiv preprint arXiv:2306.11730}, 2023.

\bibitem{SAM-SZU}
Y.~Huang, X.~Yang, L.~Liu, H.~Zhou, A.~Chang, X.~Zhou, R.~Chen, J.~Yu, J.~Chen,
  C.~Chen, H.~Chi, X.~Hu, D.-P. Fan, F.~Dong, and D.~Ni, ``Segment anything
  model for medical images?'' \emph{arXiv preprint arXiv:2304.14660}, 2023.

\bibitem{SAM-ConcealedScenes}
G.-P. Ji, D.-P. Fan, P.~Xu, M.-M. Cheng, B.~Zhou, and L.~V. Gool, ``Sam
  struggles in concealed scenes - empirical study on "segment anything",''
  \emph{arXiv preprint arXiv:2304.06022}, 2023.

\bibitem{SAM-Realworld}
W.~Ji, J.~Li, Q.~Bi, W.~Li, and L.~Cheng, ``Segment anything is not always
  perfect: An investigation of sam on different real-world applications,''
  \emph{arXiv preprint arXiv:2304.05750}, 2023.

\bibitem{skinSAM}
M.~Hu, Y.~Li, and X.~Yang, ``Skinsam: Empowering skin cancer segmentation with
  segment anything model,'' \emph{arXiv preprint arXiv:2304.13973}, 2023.

\bibitem{PolypSAM}
Y.~Li, M.~Hu, and X.~Yang, ``Polyp-sam: Transfer sam for polyp segmentation,''
  \emph{arXiv preprint arXiv:2305.00293}, 2023.

\bibitem{MedSAM}
J.~Ma and B.~Wang, ``Segment anything in medical images,'' \emph{arXiv preprint
  arXiv:2304.12306}, 2023.

\bibitem{Med-SAM-Adapter}
J.~Wu, R.~Fu, H.~Fang, Y.~Liu, Z.~Wang, Y.~Xu, Y.~Jin, and T.~Arbel, ``Medical
  sam adapter: Adapting segment anything model for medical image
  segmentation,'' \emph{arXiv preprint arXiv:2304.12620}, 2023.

\bibitem{Lora}
E.~J. Hu, Y.~Shen, P.~Wallis, Z.~Allen-Zhu, Y.~Li, S.~Wang, L.~Wang, and
  W.~Chen, ``Lora: Low-rank adaptation of large language models,'' \emph{arXiv
  preprint arXiv:2106.09685}, 2021.

\bibitem{SAMed}
K.~Zhang and D.~Liu, ``Customized segment anything model for medical image
  segmentation,'' \emph{arXiv preprint arXiv:2304.13785}, 2023.

\bibitem{3DSAM-adapter}
S.~Gong, Y.~Zhong, W.~Ma, J.~Li, Z.~Wang, J.~Zhang, P.-A. Heng, and Q.~Dou,
  ``3dsam-adapter: Holistic adaptation of sam from 2d to 3d for promptable
  medical image segmentation,'' \emph{arXiv preprint arXiv:2306.13465}, 2023.

\bibitem{SAM-Path}
J.~Zhang, K.~Ma, S.~Kapse, J.~Saltz, M.~Vakalopoulou, P.~Prasanna, and
  D.~Samaras, ``Sam-path: A segment anything model for semantic segmentation in
  digital pathology,'' \emph{arXiv preprint arXiv:2307.09570}, 2023.

\bibitem{chai2023ladder}
S.~Chai, R.~K. Jain, S.~Teng, J.~Liu, Y.~Li, T.~Tateyama, and Y.-w. Chen,
  ``Ladder fine-tuning approach for sam integrating complementary network,''
  \emph{arXiv preprint arXiv:2306.12737}, 2023.

\bibitem{SAM-3DSlicer}
Y.~Liu, J.~Zhang, Z.~She, A.~Kheradmand, and M.~Armand, ``Samm (segment any
  medical model): A 3d slicer integration to sam,'' \emph{arXiv preprint
  arXiv:2304.05622}, 2023.

\bibitem{3Dslicer}
A.~Fedorov, R.~R. Beichel, J.~Kalpathy-Cramer, J.~Finet, J.-C. Fillion-Robin,
  S.~Pujol, C.~Bauer, D.~L. Jennings, F.~M. Fennessy, M.~Sonka, J.~M. Buatti,
  S.~R. Aylward, J.~V. Miller, S.~D. Pieper, and R.~Kikinis, ``3d slicer as an
  image computing platform for the quantitative imaging network.''
  \emph{Magnetic resonance imaging}, vol. 30 9, pp. 1323--41, 2012.

\bibitem{All-in-SAM}
C.~Cui, R.~Deng, Q.~Liu, T.~Yao, S.~Bao, L.~W. Remedios, Y.~Tang, and Y.~Huo,
  ``All-in-sam: from weak annotation to pixel-wise nuclei segmentation with
  prompt-based finetuning,'' \emph{arXiv preprint arXiv:2307.00290}, 2023.

\bibitem{MedLSAM}
W.~Lei, X.~Wei, X.~Zhang, K.~Li, and S.~Zhang, ``Medlsam: Localize and segment
  anything model for 3d medical images,'' \emph{arXiv preprint
  arXiv:2306.14752}, 2023.

\bibitem{DeSAM}
Y.~Gao, W.~Xia, D.~Hu, and X.~Gao, ``Desam: Decoupling segment anything model
  for generalizable medical image segmentation,'' \emph{arXiv preprint
  arXiv:2306.00499}, 2023.

\bibitem{SAM-U}
G.~Deng, K.~Zou, K.~Ren, M.~Wang, X.~Yuan, S.~Ying, and H.~Fu, ``Sam-u:
  Multi-box prompts triggered uncertainty estimation for reliable sam in
  medical image,'' \emph{arXiv preprint arXiv:2307.04973}, 2023.

\bibitem{IA-SAM}
Y.~Zhang, T.~Zhou, P.~Liang, and D.~Z. Chen, ``Input augmentation with sam:
  Boosting medical image segmentation with segmentation foundation model,''
  \emph{arXiv preprint arXiv:2304.11332}, 2023.

\bibitem{qu2023annotating}
C.~Qu, T.~Zhang, H.~Qiao, J.~Liu, Y.~Tang, A.~L. Yuille, and Z.~Zhou, ``Segment
  anything,'' \emph{arXiv preprint arXiv:2305.09666}, 2023.

\bibitem{moor2023foundation}
M.~Moor, O.~Banerjee, Z.~F.~H. Abad, H.~M. Krumholz, J.~Leskovec, E.~J. Topol,
  and P.~Rajpurkar, ``Foundation models for generalist medical artificial
  intelligence,'' \emph{Nature}, vol. 616, pp. 259--265, 2023.

\bibitem{willemink2022toward}
M.~J. Willemink, H.~R. Roth, and V.~Sandfort, ``Toward foundational deep
  learning models for medical imaging in the new era of transformer networks.''
  \emph{Radiology. Artificial intelligence}, vol.~46, p. 210284, 2022.

\bibitem{2-5D}
Y.~Zhang, Q.~Liao, L.~Ding, and J.~Zhang, ``Bridging 2d and 3d segmentation
  networks for computation-efficient volumetric medical image segmentation: An
  empirical study of 2.5 d solutions,'' \emph{Computerized Medical Imaging and
  Graphics}, p. 102088, 2022.

\bibitem{SemiSurvey}
R.~Jiao, Y.~Zhang, L.~Ding, R.~Cai, and J.~Zhang, ``Learning with limited
  annotations: A survey on deep semi-supervised learning for medical image
  segmentation,'' \emph{arXiv preprint arXiv:2207.14191}, 2022.

\bibitem{MIA-Imperfect}
N.~Tajbakhsh, L.~Jeyaseelan, Q.~Li, J.~N. Chiang, Z.~Wu, and X.~Ding,
  ``Embracing imperfect datasets: A review of deep learning solutions for
  medical image segmentation,'' \emph{Medical Image Analysis}, vol.~63, p.
  101693, 2020.

\bibitem{SAM--Med}
C.~Wang, D.~Li, S.~Wang, C.~Zhang, Y.~Wang, Y.~Liu, and G.~Yang, ``Sam-med: A
  medical image annotation framework based on large vision model,'' \emph{arXiv
  preprint arXiv:2307.05617}, 2023.

\bibitem{huang2023push}
Z.~Huang, H.~Liu, H.~Zhang, F.~Xing, A.~Laine, E.~Angelini, C.~Hendon, and
  Y.~Gan, ``Push the boundary of sam: A pseudo-label correction framework for
  medical segmentation,'' \emph{arXiv preprint arXiv:2308.00883}, 2023.

\bibitem{Scribbleseg-1}
X.~Luo, M.~Hu, W.~Liao, S.~Zhai, T.~Song, G.~Wang, and S.~Zhang,
  ``Scribble-supervised medical image segmentation via dual-branch network and
  dynamically mixed pseudo labels supervision,'' in \emph{International
  Conference on Medical Image Computing and Computer-Assisted
  Intervention}.\hskip 1em plus 0.5em minus 0.4em\relax Springer, 2022, pp.
  528--538.

\bibitem{Scribbleseg-2}
K.~Zhang and X.~Zhuang, ``Cyclemix: A holistic strategy for medical image
  segmentation from scribble supervision,'' in \emph{Proceedings of the
  IEEE/CVF Conference on Computer Vision and Pattern Recognition}, 2022, pp.
  11\,656--11\,665.

\bibitem{GazeSAM}
B.~Wang, A.~Aboah, Z.~Zhang, and U.~Bagci, ``Gazesam: What you see is what you
  segment,'' \emph{arXiv preprint arXiv:2304.13844}, 2023.

\bibitem{SAM-UIG}
G.~Ning, H.~Liang, Z.~Jiang, H.~Zhang, and H.~Liao, ``The potential of 'segment
  anything' (sam) for universal intelligent ultrasound image guidance.''
  \emph{Bioscience trends}, 2023.

\bibitem{lappas2022interobserver}
G.~Lappas, N.~Staut, N.~G. Lieuwes, R.~Biemans, C.~J. Wolfs, S.~J. van Hoof,
  L.~J. Dubois, and F.~Verhaegen, ``Inter-observer variability of organ
  contouring for preclinical studies with cone beam computed tomography
  imaging,'' \emph{Physics and Imaging in Radiation Oncology}, vol.~21, pp. 11
  -- 17, 2022.

\end{thebibliography}

%\begin{thebibliography}{1}

%H.~Kopka and P.~W. Daly, \emph{A Guide to \LaTeX}, %3rd~ed.\hskip 1em plus

%\end{thebibliography}


%\end{IEEEbiography}


%\newpage



%\vfill


\end{document}